\documentclass[a4paper]{jpconf}
\usepackage{graphicx}
\usepackage{epsfig}
\usepackage{lineno}
\newcommand{\snnbf} {\ensuremath{\mathbf{{\sqrt{s_{\mathbf NN}}}}}}

\begin{document}

%----------------------
\title{Multi-strange baryon production in \mbox{Pb--Pb} and pp collisions at \snnbf\ = 2.76 TeV with the ALICE experiment at the LHC}
\author{Domenico Colella (for the ALICE Collaboration)}
\address{Dipartimento Interateneo di Fisica ''M. Merlin'' and Sezione INFN, Via Orabona 4, 70126 Bari, Italy}
\ead{domenico.colella@cern.ch}

%--------------------
\begin{abstract}
The production of $\Xi^{-}$ and $\Omega^{-}$ baryons and their anti-particles in \mbox{Pb--Pb} and pp collisions at $\sqrt{s_{\rm NN}}$ = 2.76 TeV has been measured by the ALICE collaboration. The transverse momentum spectra at mid-rapidity ($|y| < 0.5$) in pp and in \mbox{Pb--Pb} collisions for five centrality intervals have been compared with model predictions. Hyperon yields and spectra in \mbox{Pb--Pb}  collisions, normalized to the corresponding measurements in pp at the same centre-of-mass energy, allow the study of the strangeness enhancement and the nuclear modification factor as a function of the transverse momentum ($p_{\rm T}$) and collision centrality.
\end{abstract}

%---------------------------
\section{Introduction}
The study of strange and multi-strange particle production in relativistic heavy-ion (\mbox{A--A}) interactions is an important tool to investigate the properties of the strongly interacting system created in the collision, as there is no net strangeness content in the colliding nuclei. In particular, baryons with more than one unit of strangeness are very useful probes of the early partonic stages of the collision due to their small hadronic cross-section.

The enhancement of strangeness in heavy-ion collisions was one of the earliest proposed signals for the Quark-Gluon Plasma \cite{Rafelski}. It rests on the expectation that in a deconfined state the abundances of parton species should quickly reach their equilibrium values, resulting in a higher abundance of strangeness per participant than that seen in pp interactions. 
This enhancement, first observed at SPS \cite{NA57_2010} and confirmed afterwards by data at RHIC \cite{STAR}, is more pronounced for particles with larger strangeness content and decreases as the centre-of-mass energy of the collision increases.
Over the past 15 years, it has been found that the canonical suppression effect, for pp collisions, is important \cite{Redlich_2} and contributes to the overall hyperon enhancement in \mbox{A--A}. This interpretation can now be re-examined at the much higher LHC energy. 

A suppression of high-momentum unidentified particle production in \mbox{Pb--Pb} compared to pp collisions has been observed by the ALICE collaboration and used as evidence for strong parton energy loss and large medium density at the LHC \cite{ALI11}. Measurements of the nuclear modification factors for identified particles could provide insight into particle production and energy loss mechanisms. 

The ALICE experiment was specifically designed to study heavy-ion collisions at the LHC, namely the properties of strongly interacting matter at high energy density. The first LHC heavy-ion run took place at the end of 2010 with \mbox{Pb--Pb} ions accelerated at a centre-of-mass energy $\sqrt{s_{NN}}$ = 2.76 TeV. Almost $30\times10^{6}$ minimum bias nuclear interaction triggers were recorded. The pp reference data were collected during March 2011. Almost $80\times10^{6}$ minimum bias nucleon interaction triggers were recorded. 
The tracking and vertexing are performed with the full tracking system: the Inner Tracking System (ITS, six layers of silicon detectors) and the Time Projection Chamber (TPC), which is also used for particle identification via specific ionization. The Silicon Pixel Detector (SPD, the two innermost layers of the ITS) and the VZERO detector (scintillation hodoscopes placed on either side of the interaction point) were used for triggering. The VZERO was also crucial for the collision centrality determination. A complete description of the ALICE sub-detectors can be found in \cite{JINST}.

%---------------------------
\section{Analysis method}
Multi-strange baryons are reconstructed through their weak decay topologies, namely $\Xi^{-} \rightarrow \pi^{-} + \Lambda$, $\Omega^{-} \rightarrow \rm{K^{-}} + \Lambda$ (with $\Lambda \rightarrow\pi^{-}  + \rm p$), and the corresponding charge conjugate decays for the anti-particles.
The resulting branching ratios are 63.9\% and 43.3\%, for the $\Xi$ and the $\Omega$ respectively.
The $\Xi$ and the $\Omega$ candidates are found by combining reconstructed charged tracks: cuts on geometry and kinematics are applied to select first the $\Lambda$ candidate and then to match it with all the remaining secondary tracks (bachelor candidates).
In addition, cuts on particle identification via specific energy loss in the TPC for the three daughter tracks are used. 

To extract the signal in $p_{\rm T}$ intervals, a symmetric region around the invariant mass peak ($\pm$3$\sigma$) is defined by fitting the distribution with the sum of a Gaussian and a polynomial.
The background is sampled in two regions on both sides of the peak and fitted with a polynomial of first or second degree (depending on the colliding system and the $p_{\rm T}$ interval). In each $p_{\rm T}$ interval the signal in the peak region is obtained by subtracting the integral of the background fit function from the peak population. 
More details can be found in \cite{PbPbPaper}.

%---------------------------
\section{Results and conclusions}
\begin{figure}[h]
\centering 
\includegraphics[width=18pc]{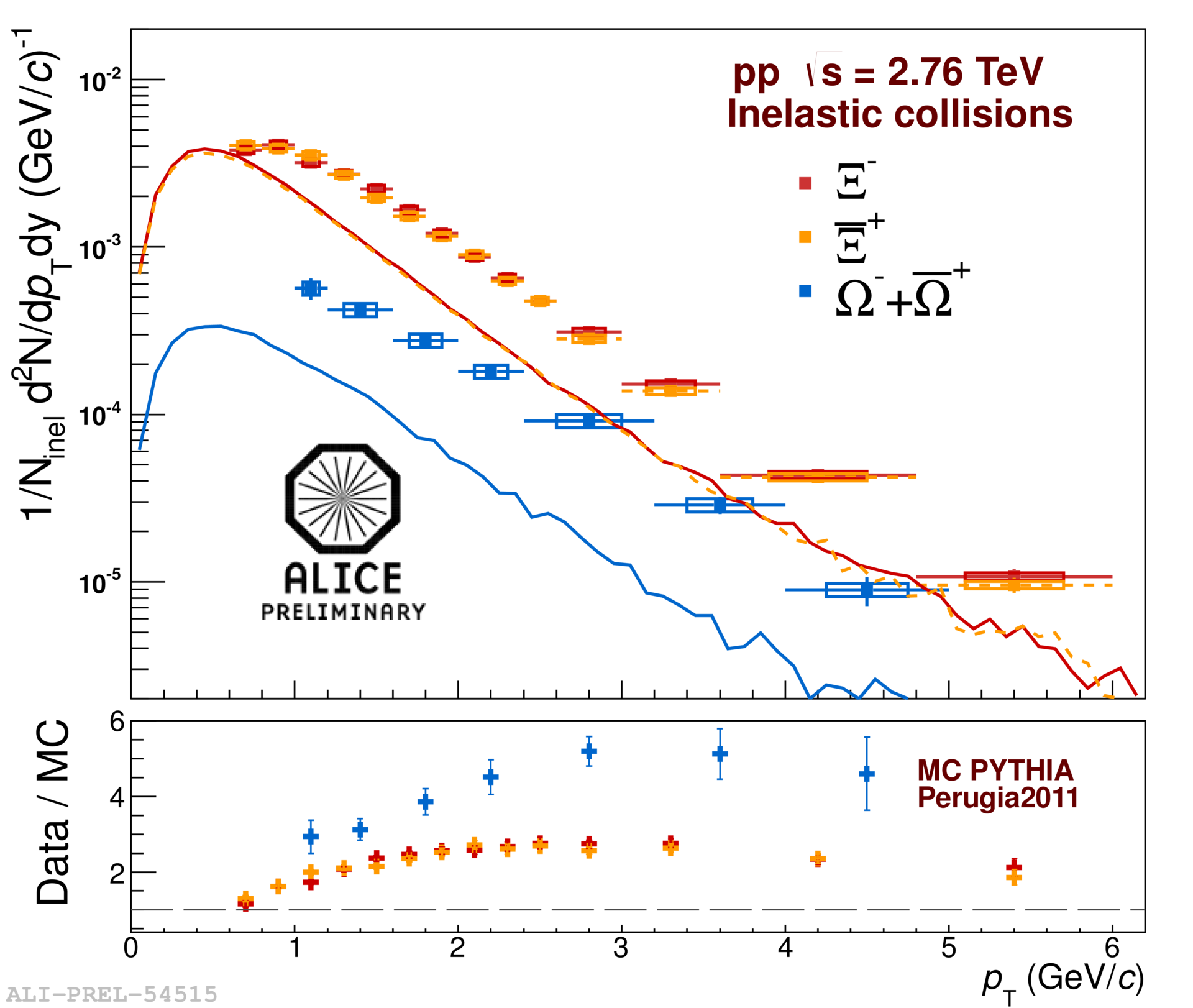}%
\caption{\label{Fig:pp} Transverse momentum spectra normalized to the number of inelastic collisions for $\Xi^{-}$, $\overline{\Xi}^{+}$ and $\Omega$ in pp collisions, compared with predictions obtained with PYTHIA 6 (Perugia-2011 tune \cite{pythia}\cite{pythiaarXiv}). Ratios of data to models are also shown. The error bars represent the statistical uncertainties and the boxes the systematic ones.}
\end{figure}

\begin{figure}[h]
\centering
\includegraphics[width=14pc]{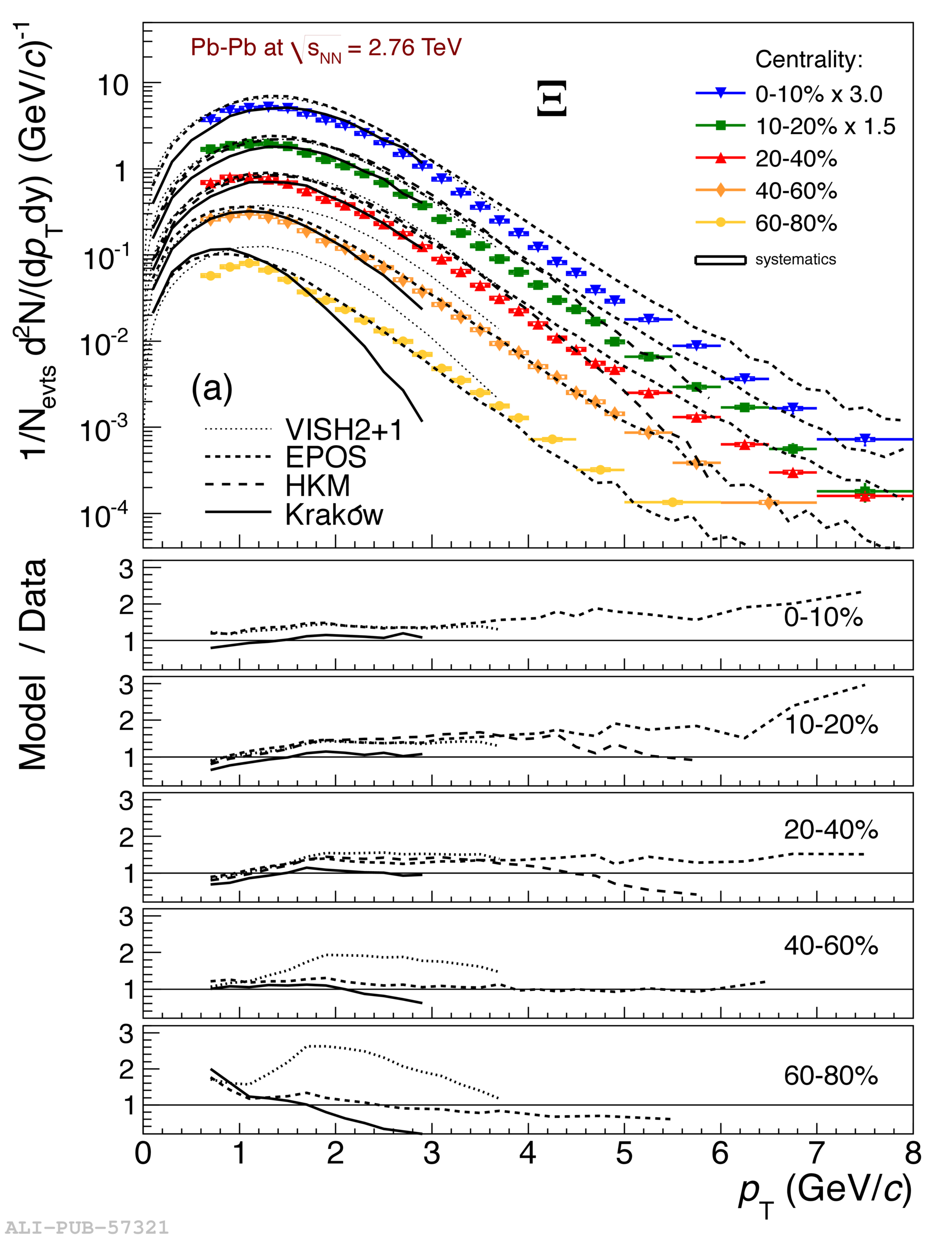}
\includegraphics[width=14pc]{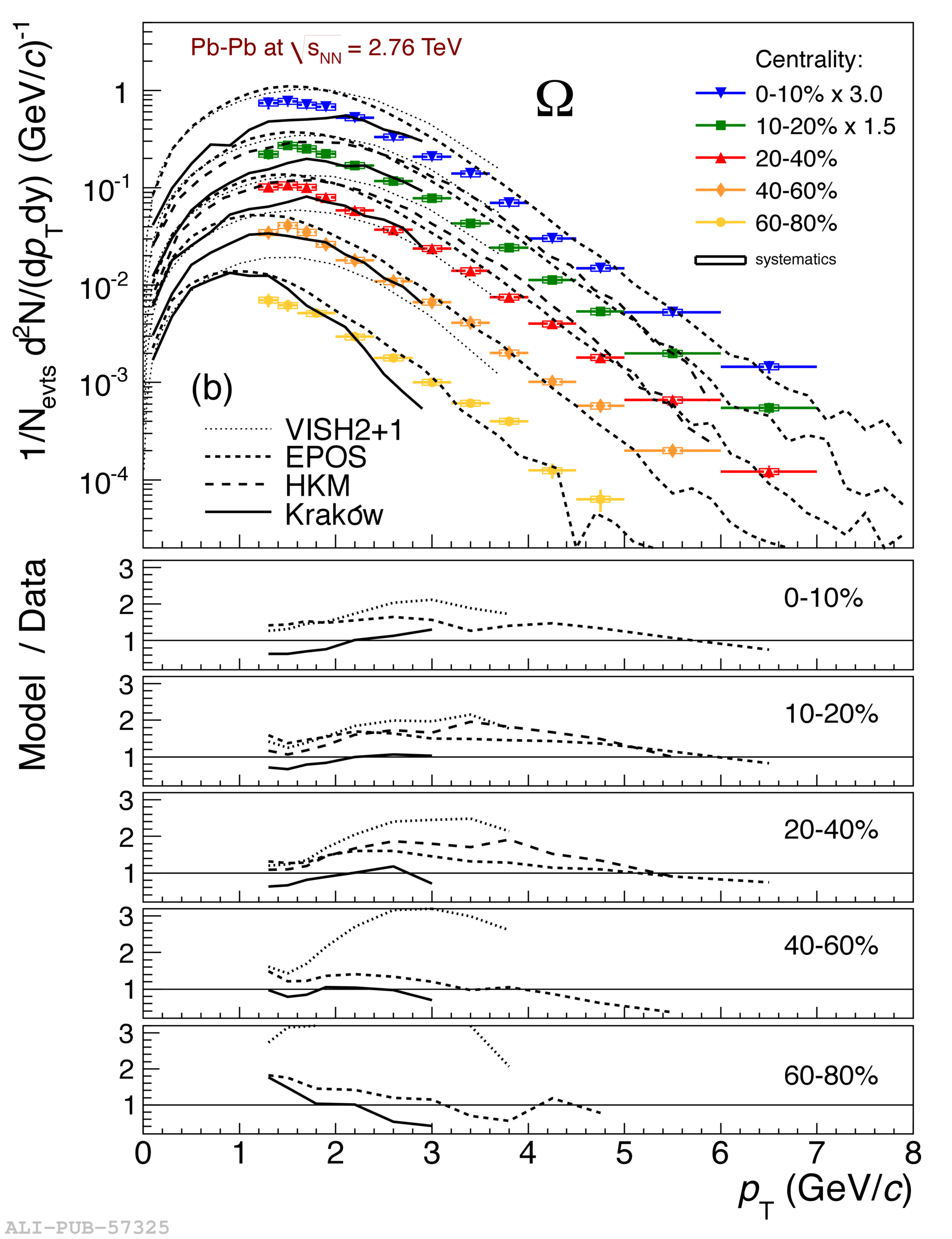}
\caption{\label{Fig:PbPb} Transverse momentum spectra for $\Xi$ (a) and $\Omega$ (b) hyperons (average of particle and anti-particle) in five different centrality classes, compared to hydrodynamic models. Ratios of models to data are also shown.}
\end{figure}

Transverse momentum spectra, corrected for acceptance and efficiency, in pp and \mbox{Pb--Pb} collisions are shown in Figure \ref{Fig:pp} and \ref{Fig:PbPb} respectively. In pp collisions, the comparison with predictions obtained using PYTHIA 6 (Perugia-2011 tune \cite{pythia}\cite{pythiaarXiv}) shows that this model underestimates the yields, both for  $\Xi$ and $\Omega$, and does not reproduce the shape of the spectra. Spectra in \mbox{Pb--Pb} collisions, reconstructed in five centrality classes (0-10\%, 10-20\%, 20-40\%, 40-60\% and 60-80\%), are compared with predictions obtained using four hydrodynamic models: VISH2+1 \cite{vish2+1}
, HKM \cite{HKM}
, Krak\'ow \cite{krakow} 
and EPOS  \cite{EPOS}
. Krak\'ow provides a good description for both yields and shapes in the $p_{\rm T}$ range up to 3 GeV/\textit{c}, and EPOS gives the most successful description of  spectral shape in a wider $p_{\rm T}$ range. The overall description for the $\Omega$ is less successful while for both particles types it degrades going from central to more peripheral collisions.

\begin{figure}[h]
\centering
\includegraphics[width=24pc]{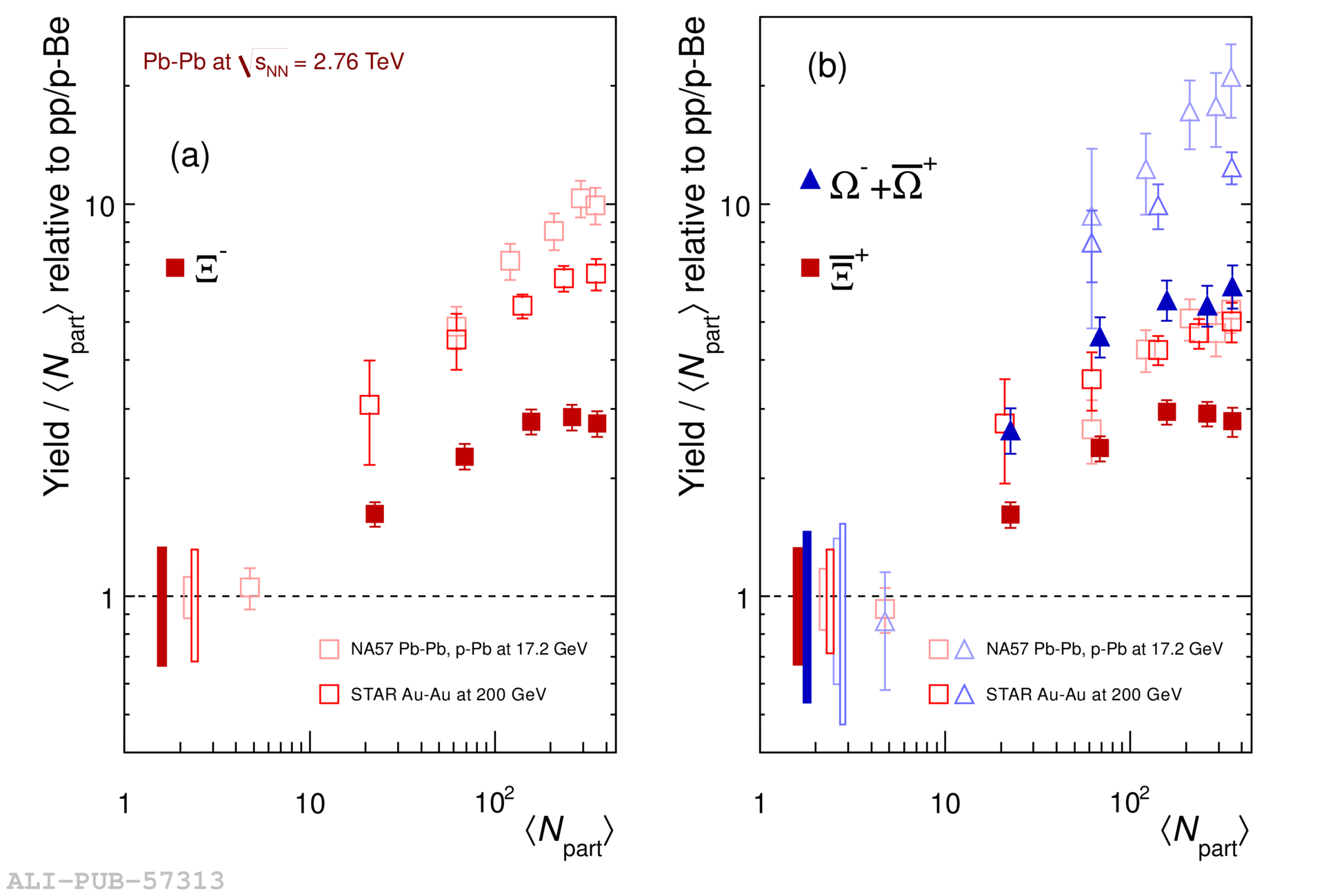}
\includegraphics[width=12pc]{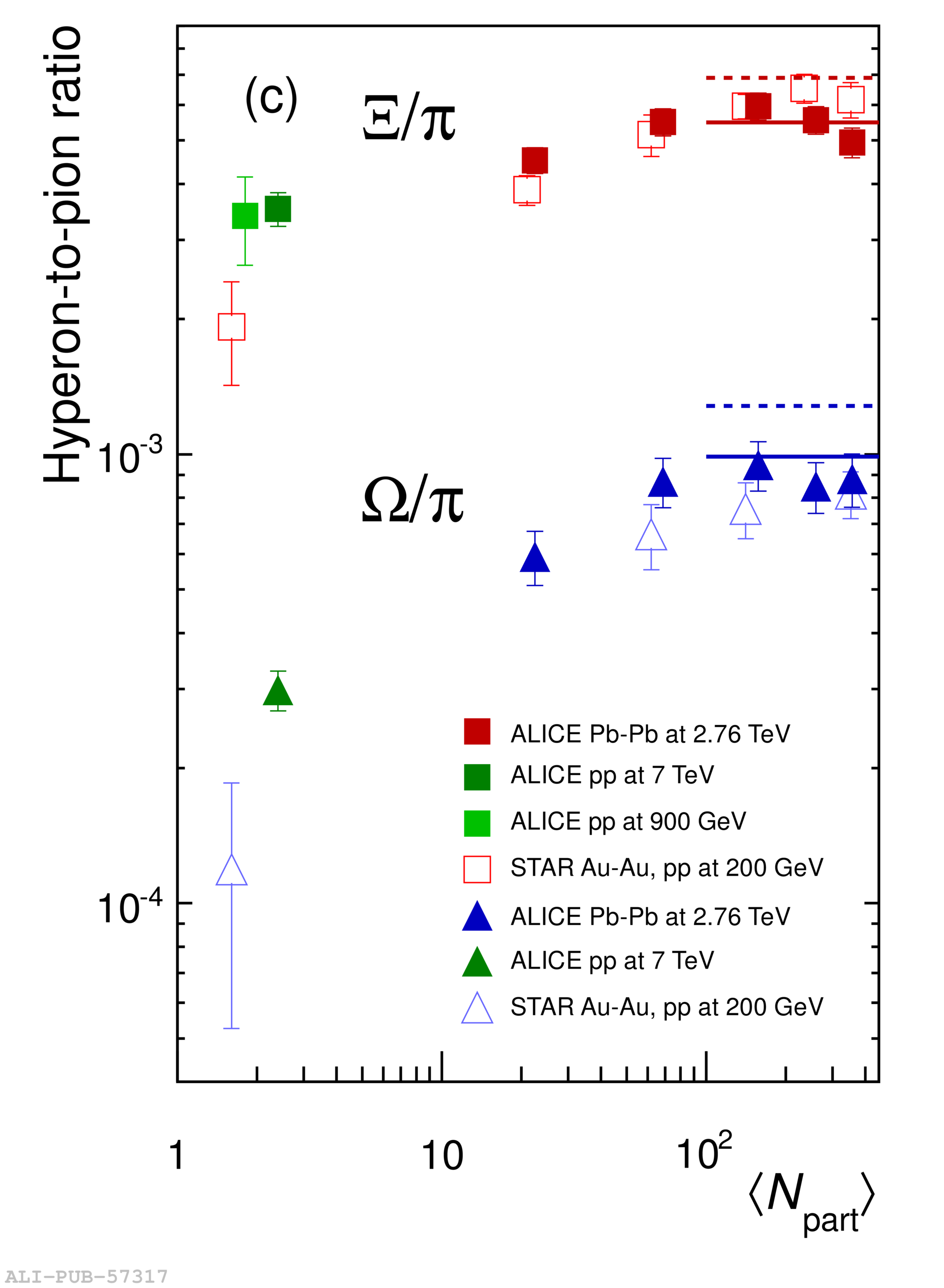}
\caption{\label{Fig:enhance} (a,b) Enhancements in the rapidity range $|y| < 0.5$ as a function of the mean number of participants $\langle{N_{\rm part}}\rangle$. Boxes on the dashed line at unity indicate statistical and systematic uncertainties on the pp or \mbox{p--Be} reference. (c) Hyperon-to-pion ratios as a function of $\langle{N_{\rm part}}\rangle$, for \mbox{A--A} and pp collisions at LHC and RHIC energies. The lines mark the thermal model predictions from \cite{thermod1} (full line) and \cite{thermod2} (dashed line).}
\end{figure}

The strangeness enhancements, calculated as the ratio between the yields in \mbox{Pb--Pb} collisions and those in pp interactions at the same energy, both normalized to the number of participants ($N_{\rm part}$), are shown in Figure \ref{Fig:enhance}(a-b). The pp reference values were obtained by interpolating ALICE and STAR data at different energies \cite{PbPbPaper} and checked with the preliminary measurements of the yields in pp collisions at the reference energy. The enhancements as a function of the mean number of participants increase with centrality and with the strangeness content of the particle as already observed at lower energies, and decrease as the centre-of-mass energy increases, continuing the trend established at lower energies (between SPS and RHIC energies).

The hyperon-to-pion ratios $\Xi/\pi\equiv{(\Xi^{-}+\overline{\Xi}^{+})/(\pi^{-}+\pi^{+})}$ and $\Omega/\pi\equiv{(\Omega^{-}+\overline{\Omega}^{+})/(\pi^{-}+\pi^{+})}$, for \mbox{A--A} and pp collisions, both at LHC and RHIC energies, are shown in Figure \ref{Fig:enhance}(c) as a function of $\langle{N_{\rm part}\rangle}$. The relative production of strangeness in pp collisions at the LHC is larger than at RHIC energy while it is almost the same at the two energies for \mbox{A--A} collisions, giving an explanation of the decrease of enhancement with increasing energy. The measurements of these ratios in \mbox{Pb--Pb} collisions are in agreement with predictions from the thermal model, based on a grand canonical approach \cite{thermod1}\cite{thermod2}. The hyperon-to-pion ratio increases when going from pp to \mbox{A--A}, showing a relative enhancement of strangeness production in \mbox{A--A} collisions (normalized to the pion yield) which is about half of that seen when normalizing to the mean number of participants. The enhancement rises with centrality up to about $\langle{N_{\rm part}\rangle} \sim{150}$ and apparently saturates thereafter.

The nuclear modification factor ($R_{\rm AA}$) is defined as the ratio between the differential yield in \mbox{A--A} collisions ($\textrm{d}^{2}N^{\rm AA}/\textrm{d}p_{\rm T}dy$) and the differential cross-section of particle production in pp collisions ($\textrm{d}^{2}\sigma^{\rm NN}/\textrm{d}p_{\rm T}dy$), normalized by the geometric nuclear overlap function $T_{\rm AA}$ \cite{Centrality}. 
It has been measured by ALICE in different centrality intervals and both for multi-strange baryons and lighter particles \cite{Raa} \cite{Raa2}.
The $R_{\rm AA}$ for $\Xi$ follows the same trend as the proton at high $p_{\rm T}$ ($>$ 5 GeV/\textit{c}), where the suppression does not depend on the particle mass. At intermediate $p_{\rm T}$ there are indications of mass-ordering among the baryons. The $R_{\rm AA}$ for the $\Omega$ seems to be strongly affected by the strangeness enhancement and the effect becomes weaker when going from the most central to the most peripheral class. 

%---------------------------

\end{document}